\documentclass[pra,showpacs, twocolumn,floatfix,superscriptaddress]{revtex4-1}
\usepackage[colorlinks=true,linkcolor=blue,citecolor=blue,urlcolor=blue]{hyperref}
\usepackage{graphicx}
\usepackage{bm}
\usepackage{amsmath}
\usepackage[squaren,Gray]{SIunits} 
\usepackage{natbib}
\usepackage{url}
\usepackage{textcase}

\providecommand{\ket}[1]{\lvert #1 \rangle}

\newcommand{\ud}{\mathrm{d}}
\usepackage{mathbbol}

\usepackage{xcolor}

\usepackage{subfig}

\newcommand{\modif}{}

\begin{document}

\title{Toolbox for continuous variable entanglement production and measurement using  spontaneous parametric down conversion}

\author{G.~Boucher}
\thanks{These authors have equally contributed to this work.}
\affiliation{Laboratoire Mat\'eriaux et Ph\'enom\`enes Quantiques, Universit\'e Paris Diderot, CNRS UMR 7162, 75013, Paris, France}

\author{T.~Douce}
\thanks{These authors have equally contributed to this work.}
\affiliation{Laboratoire Mat\'eriaux et Ph\'enom\`enes Quantiques, Universit\'e Paris Diderot, CNRS UMR 7162, 75013, Paris, France}

\author{D.~Bresteau}
\altaffiliation{Current adress: Laboratoire Aim\'e Cotton, B\^atiment 505--Campus d'Orsay, Univ. Paris-Sud 11, 91405 Orsay Cedex, France}
\affiliation{Laboratoire Mat\'eriaux et Ph\'enom\`enes Quantiques, Universit\'e Paris Diderot, CNRS UMR 7162, 75013, Paris, France}

\author{S.~P.~Walborn}
\affiliation{Instituto de F\'{\i}sica, Universidade Federal do Rio de Janeiro. Caixa Postal 68528, 21941-972 Rio de Janeiro, RJ, Brazil}

\author{A.~Keller}
\affiliation{Institut de Sciences Mol\'eculaires d'Orsay (CNRS), Univ. Paris-Sud 11, B\^{a}timent 350--Campus d'Orsay, 91405 Orsay Cedex, France}

\author{T.~Coudreau}
\affiliation{Laboratoire Mat\'eriaux et Ph\'enom\`enes Quantiques, Universit\'e Paris Diderot, CNRS UMR 7162, 75013, Paris, France}

\author{S.~Ducci}
\affiliation{Laboratoire Mat\'eriaux et Ph\'enom\`enes Quantiques, Universit\'e Paris Diderot, CNRS UMR 7162, 75013, Paris, France}

\author{P.~Milman}
\affiliation{Laboratoire Mat\'eriaux et Ph\'enom\`enes Quantiques, Universit\'e Paris Diderot, CNRS UMR 7162, 75013, Paris, France}

\begin{abstract}
We provide a toolbox for continuous variables quantum state engineering and \modif{characterization} of biphoton states produced by spontaneous parametric down conversion in a transverse pump configuration. We show that the control of the pump beam's incidence spot and angle corresponds to phase space displacements of conjugate collective continuous variables of the biphoton. In particular, we illustrate \modif{with numerical simulations on a semiconductor device} how this technique can be used to engineer and characterize arbitrary states of the frequency and time degrees of freedom. \end{abstract}
\pacs{03.67.Bg, 42.50.Dv, 42.65.Lm, 42.65.Wi}
 
\maketitle

Spontaneous parametric down conversion (SPDC) experiments play a prominent role in the field of quantum information and communications. Single and multiple photon pairs generated through SPDC display entanglement in multiple  degrees of freedom (DOF) which are fundamentally different from each other. When combined or independently accessed, they constitute a powerful platform for experimental demonstrations of quantum protocols. Discrete DOF, such as polarization and orbital angular momentum, are currently used to implement quantum logic gates and protocols \cite{obrien2007SOptical}, test the non--local properties of quantum mechanics \cite{bell1964Peinstein, clauser1969PRLProposed,aspect1982PRLExperimentala,ursin2007NPEntanglement}, and realize quantum cryptography \cite{jennewein2000PRLQuantum,naik2000PRLEntangled,gisin2002RoMPQuantum} and teleportation \cite{bennett1993PRLTeleporting,bouwmeester1997NLExperimental}. DOF associated to observables with a continuous spectrum, like the quadratures of the electromagnetic field, potentially offer the same versatility as discrete ones in the field of quantum information. Continuous DOF in the single photon regime, such as frequency,  transverse momentum or position display a  perfect analogy with a multi-photon single mode continuous variables (CV) \cite{abouraddy2007PRAViolation,tasca2011PRAContinuous}. Consequently, they constitute an attractive platform to realize CV quantum information protocols \cite{braunstein2005RoMPQuantum} that are usually associated to the single mode multi-photon configuration. For example, an appealing aspect of using single photon's transverse coordinates in this field  is their relatively easy manipulation with readily available optical devices, such as  spatial light modulators (SLMs) and lenses \cite{tasca2011PRAContinuous}, circumventing the difficulties encountered in multi-photon CV strategies to implement non-gaussian operations, which are essential ingredients of universal quantum computation \cite{braunstein2005RoMPQuantum}. For these reasons, the study of entanglement in CV in the single photon regime is a valuable strategy to demonstrate CV-based quantum gates \cite{lloyd1999PRLQuantum},  quantum key distribution \cite{grosshans2002PRLContinuous,grosshans2003NQuantum}, error correcting codes \cite{gottesman2001PRAEncoding, farias2014Quantum}, quantum metrology protocols \cite{giovannetti2006PRLQuantum} and to study quantum to classical transitions \cite{zurek1991PTDecoherence, deleglise2008NLReconstruction}. Finally, we can mention that continuous DOF can be combined to discrete ones to implement conditional operations \cite{hor-meyll2014PRLAncilla,vernaz-gris2014PRAContinuous,anders2010PRAAncilla}.

CV entanglement in photon pairs can be generated in different DOF via SPDC; one example are the spatial transverse DOF
 of photon pairs  produced in non-linear bulk crystals. Tayloring the spatial properties of the pump beam has been previously employed to engineer\cite{monken1998PRATransfer,valencia2007PRLShaping} and detect quantum states of photon pairs with different properties : in \cite{walborn2003PRLMultimode} it was shown that controlling the transverse phase properties of the pump beam could lead to bunching or anti-bunching of a photon pair. Refs.~\cite{machado2013PRAInterferometric, hor-meyll2014PRLAncilla} demonstrate that the use of SLMs put in the arms of an interferometer can lead to the measurement of arbitrary momenta of the quantum state associated to the transverse photonic degrees of freedom. 
\par
Using frequency as a CV degree of freedom instead of the transverse coordinates presents a series of potential advantages: in general, the transverse variables of a propagating field are entangled in the two-dimensional orthogonal spatial coordinates $x$ and $y$, or equivalently, the momentum coordinates $p_x$ and $p_y$; this renders arbitrary state production and measurement more challenging, and limits the applications of such degrees of freedom in quantum information tasks, since the quantum state associated to each spatial direction is entangled and can be cross-correlated between $x$ and $y$ directions. Moreover, this spatial entanglement is jeopardized when coupling into optical fibers or waveguides, while frequency states remains robust in such devices. For all these reasons, we will illustrate our ideas focusing on the spectral degree of freedom.
Indeed, the characterization of arbitrary frequency entangled states of photon pairs draws a lot of attention as illustrated by recent works on frequency dependent intensity correlation measurements of the photon pair~\cite{eckstein2014LPRHigh}, the reconstruction of biphoton wavefunctions generated with a monochromatic pump~\cite{tischler2015Measurement} and amplitude sensitive tomography techniques in the time-energy space~\cite{avenhaus2014Time}.
\par
In the waveguided regime, frequency correlated, uncorrelated and anti-correlated photon pairs can be produced by modifying the spectral properties of the pump beam \cite{eckstein2011PRLHighly, cho2014PRLEngineering}. Nevertheless, even if the nature of correlations can be controlled, pump bandwidth modification alone cannot lead to arbitrary CV state engineering, an essential tool for CV quantum information and communications, fundamental tests of quantum mechanics and quantum metrology. 

\begin{figure*}
\centering
\begin{minipage}{0.8\columnwidth}

  \subfloat[]{\label{fig:source}\includegraphics[width=\columnwidth]{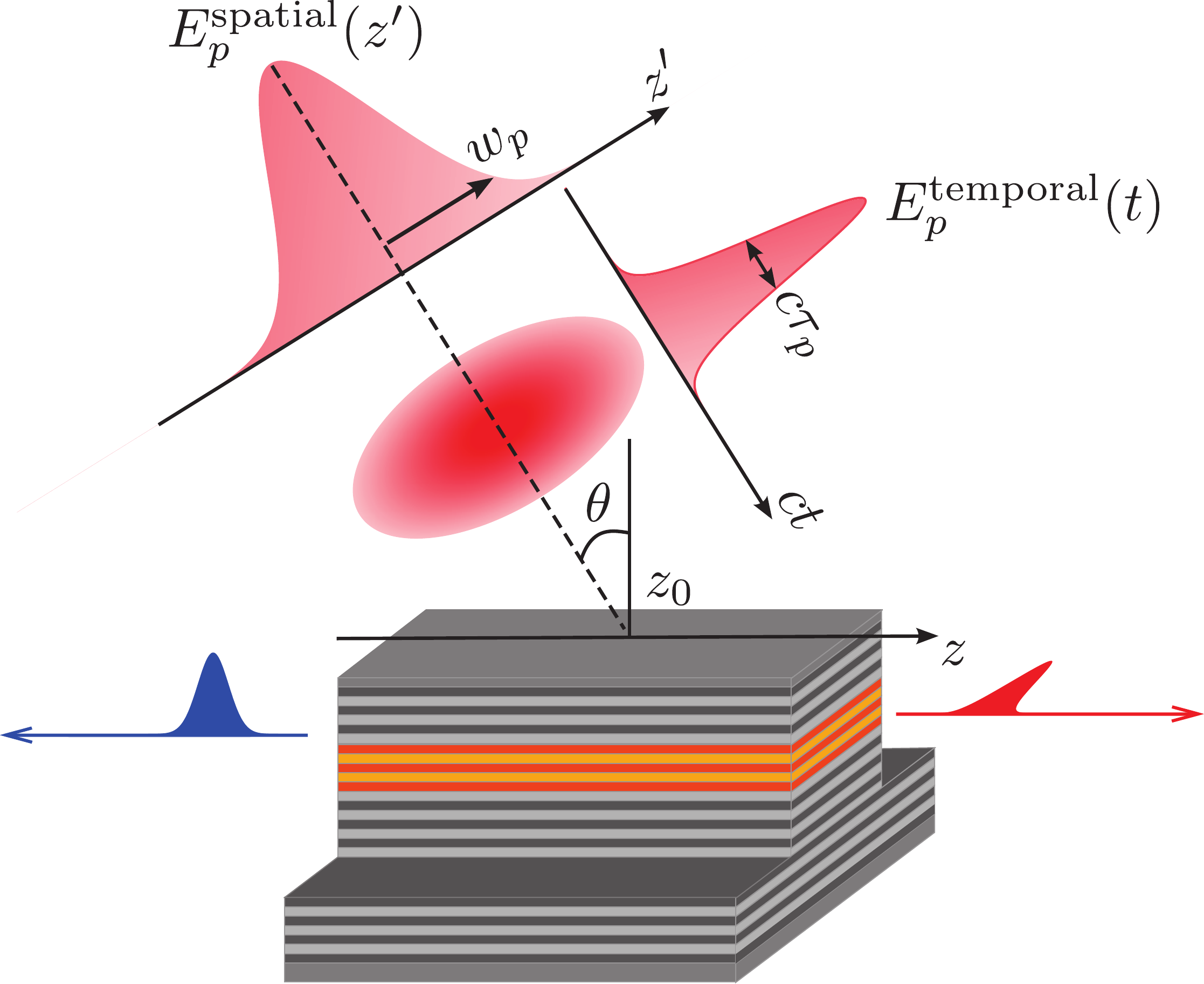}}

\end{minipage}
\begin{minipage}{0.4\columnwidth}

  \subfloat[]{\label{fig:JSA_norm}\includegraphics[width=\columnwidth]{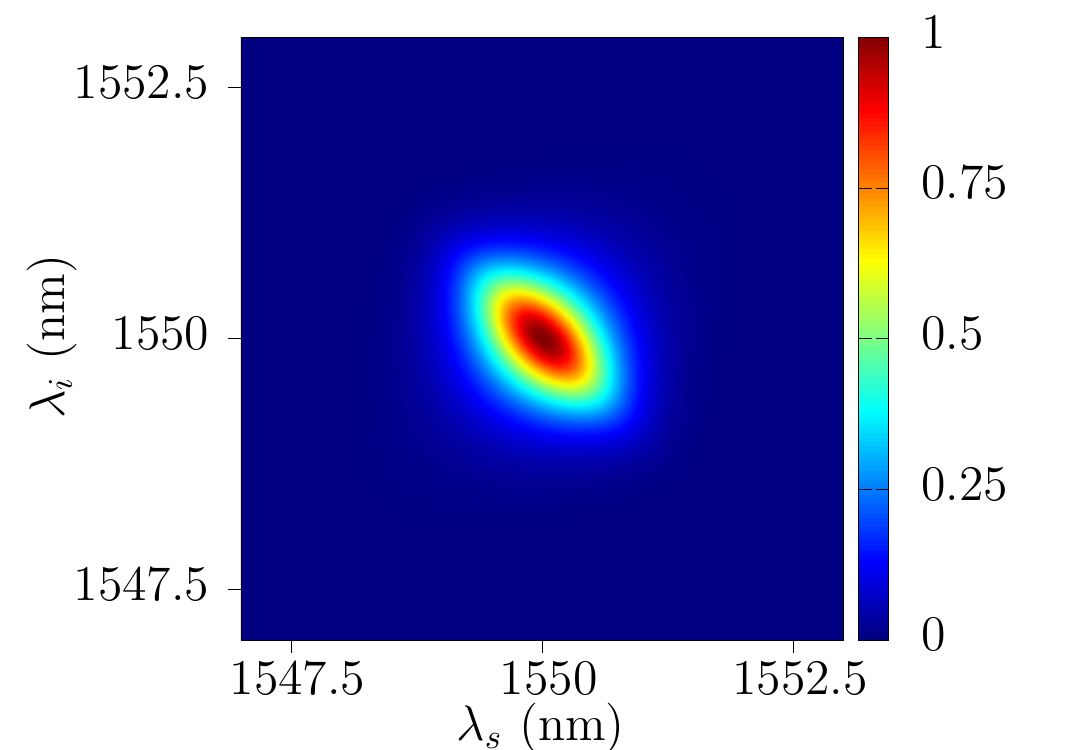}}\par
  \subfloat[]{\label{fig:JSA_phase}\includegraphics[width=\columnwidth]{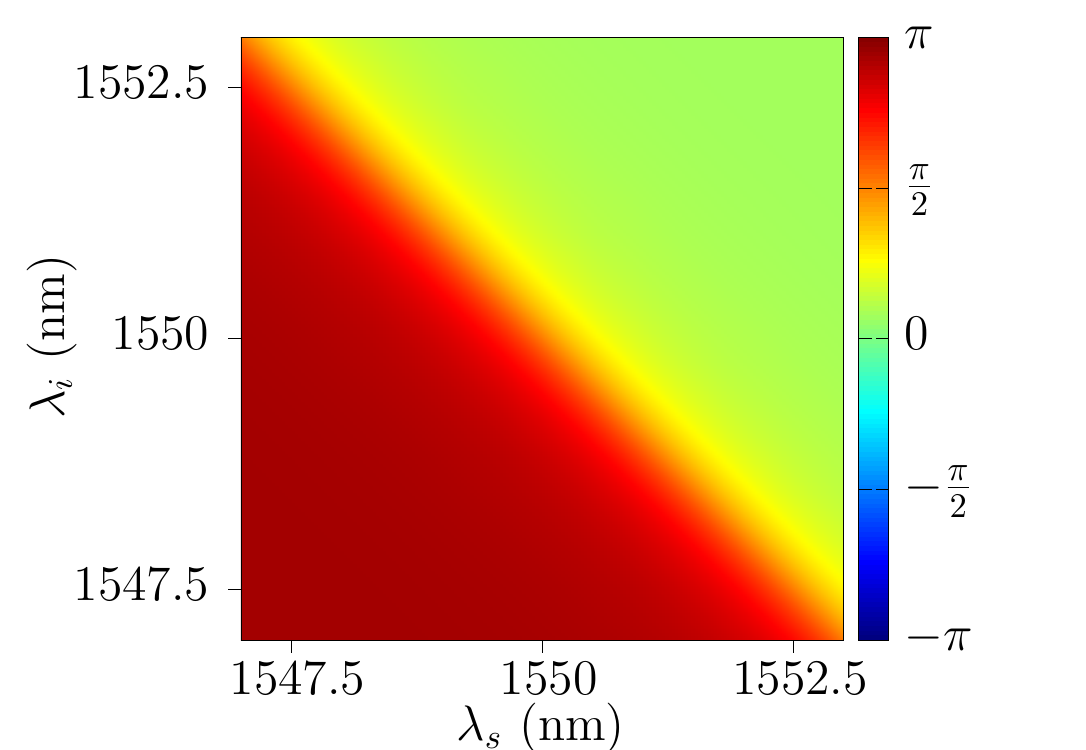}}

\end{minipage}
\begin{minipage}{0.8\columnwidth}
  \subfloat[]{\label{fig:simpleWigner}\includegraphics[width=\columnwidth]{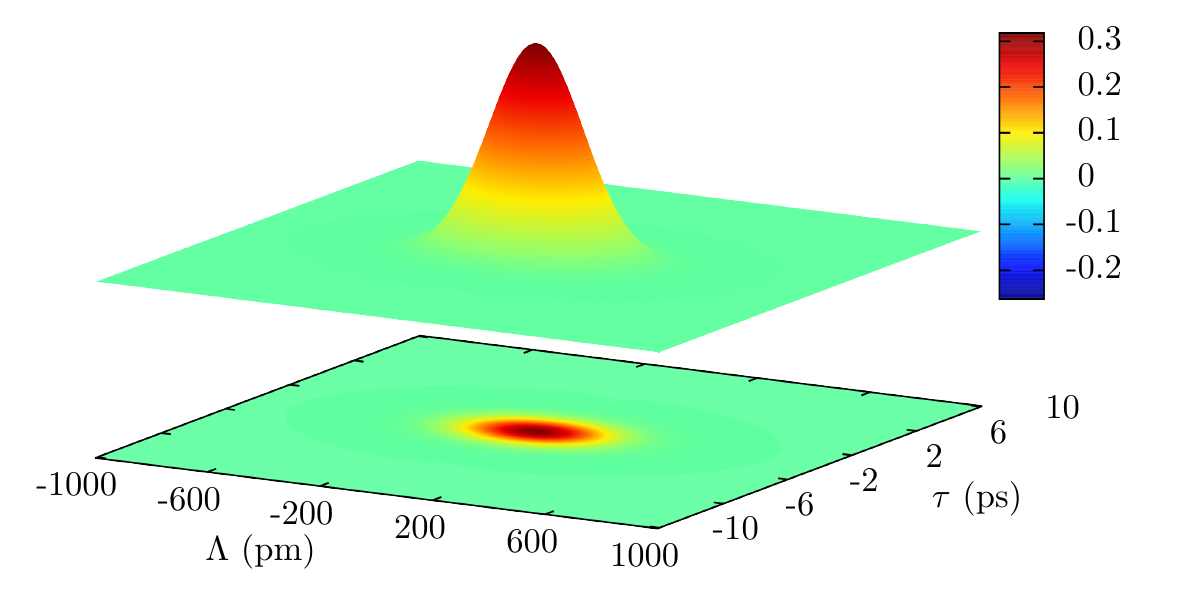}}
\end{minipage}
  \caption{(Color online) (a) Counter-propagating phase-matching scheme implemented in a semiconductor microcavity waveguide. The characteristics of the pump pulses allow to tune the time-energy properties of the biphoton. (b) Norm and (c) phase of the biphoton wavefunction for $w_p = 200\micro\meter$, $\lambda_p = 775\nano\meter$, $\tau_p = 3.2\pico\second$ and $\theta = \theta_{\mathrm{deg}}$. The phase pattern in (c) is due to the resonance of the pump in the microcavity. (d) Corresponding Wigner function. The $\Omega$ axis is given in units $\Lambda = \frac{8\pi c}{\omega_p^2}\Omega$.}
  \label{fig:simple}
\end{figure*}

In this manuscript, we provide a toolbox to exploit the capabilities of  SPDC to create, measure and characterize entanglement in the CV DOF of a photon pair. Our strategy is based on the combination of pump spatial engineering with the possibility of characterizing a Wigner function of the photon pair using a Hong-Ou-Mandel (HOM) \cite{hong1987PRLMeasurement} type experiment \cite{douce2013SRDirect}.

\begin{figure*}
\begin{minipage}{0.8\columnwidth}
\centering
  \subfloat[]{\label{fig:fourLegsCat}\includegraphics[width=\columnwidth]{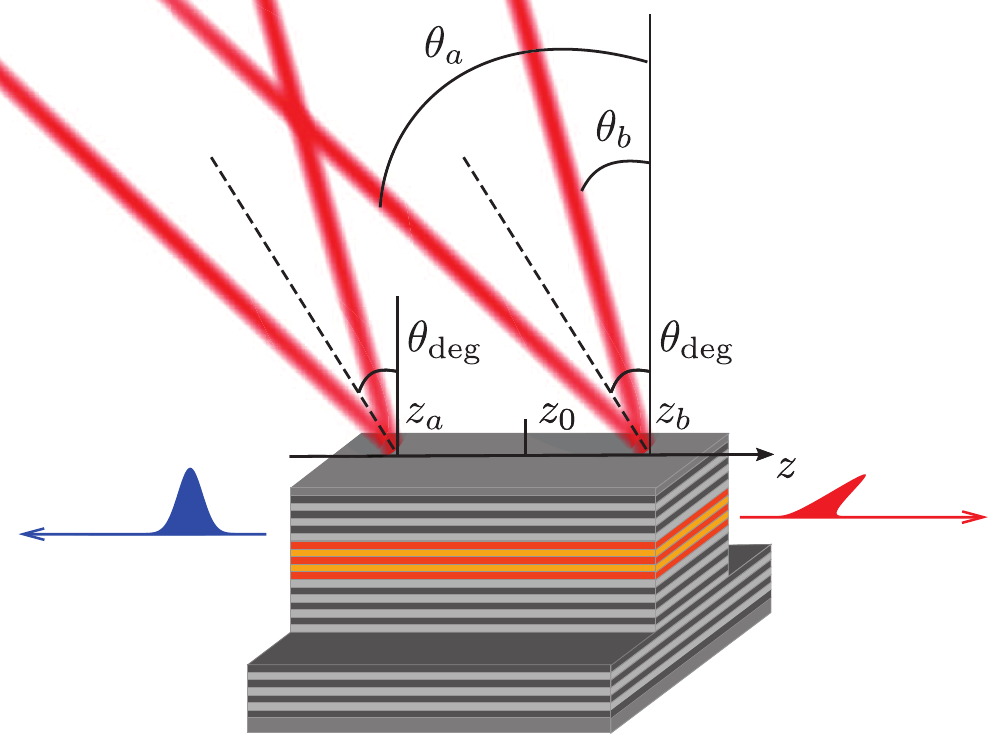}}
  \end{minipage}
  \begin{minipage}{0.4\columnwidth}

  \subfloat[]{\label{fig:Cat_norm}\includegraphics[width=\columnwidth]{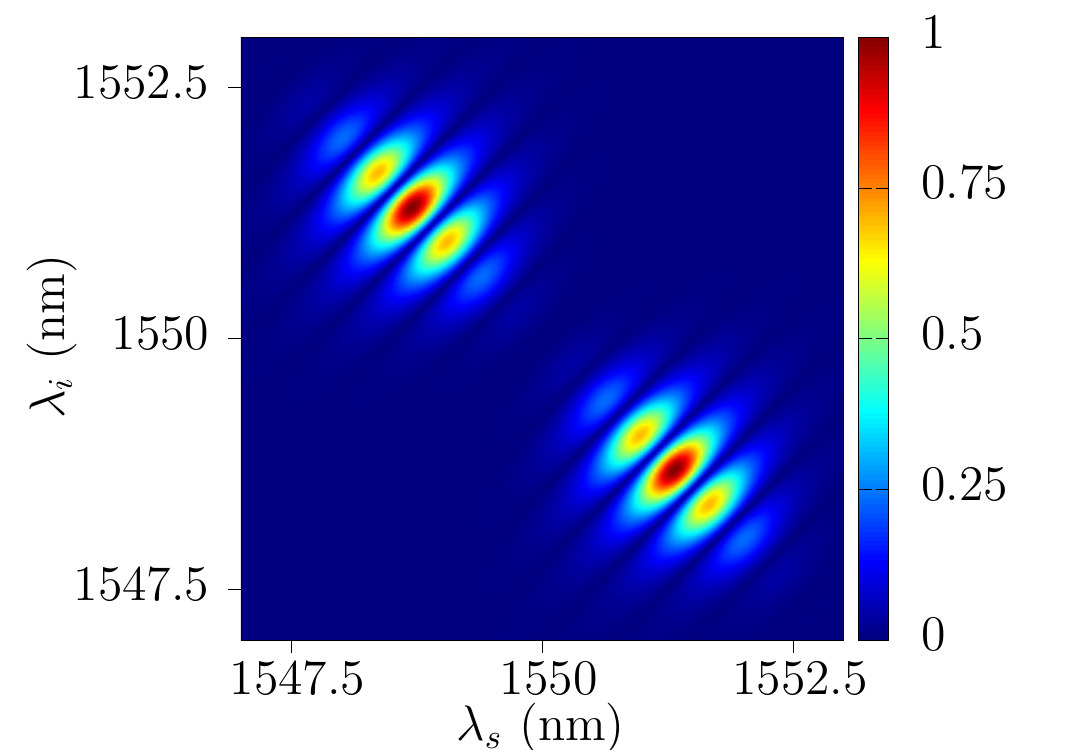}}\par
  \subfloat[]{\label{fig:Cat_phase}\includegraphics[width=\columnwidth]{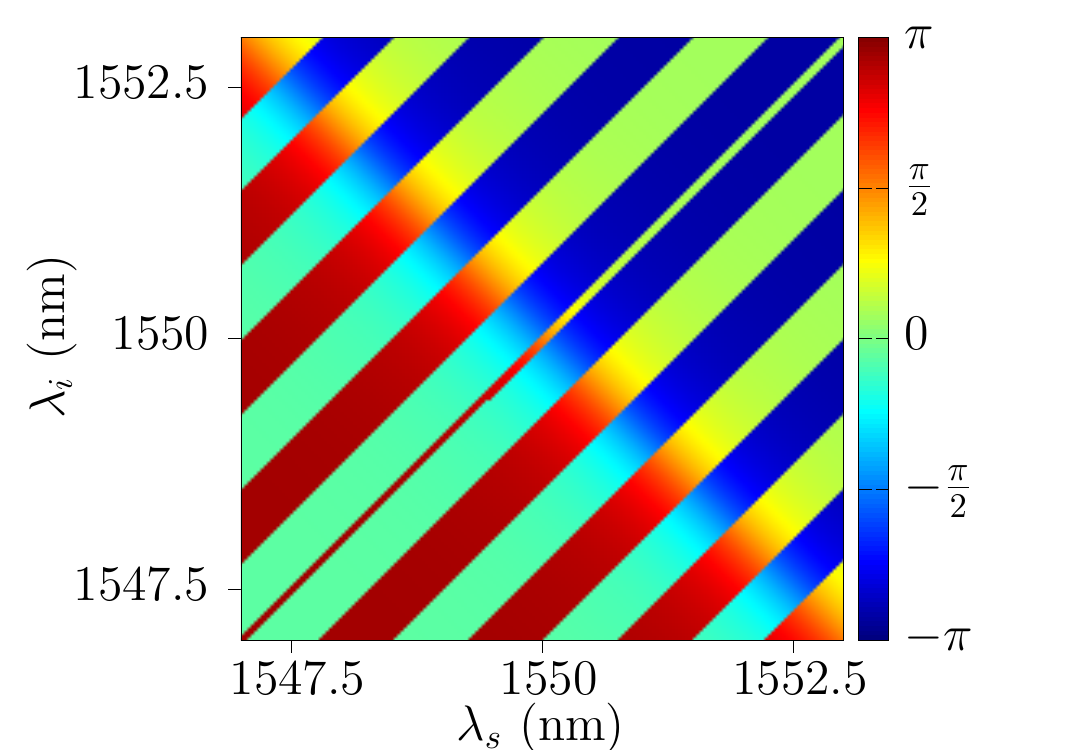}}

\end{minipage}
  \begin{minipage}{0.8\columnwidth}
\centering
  \subfloat[]{\label{fig:fourCatWigner}\includegraphics[width=\columnwidth]{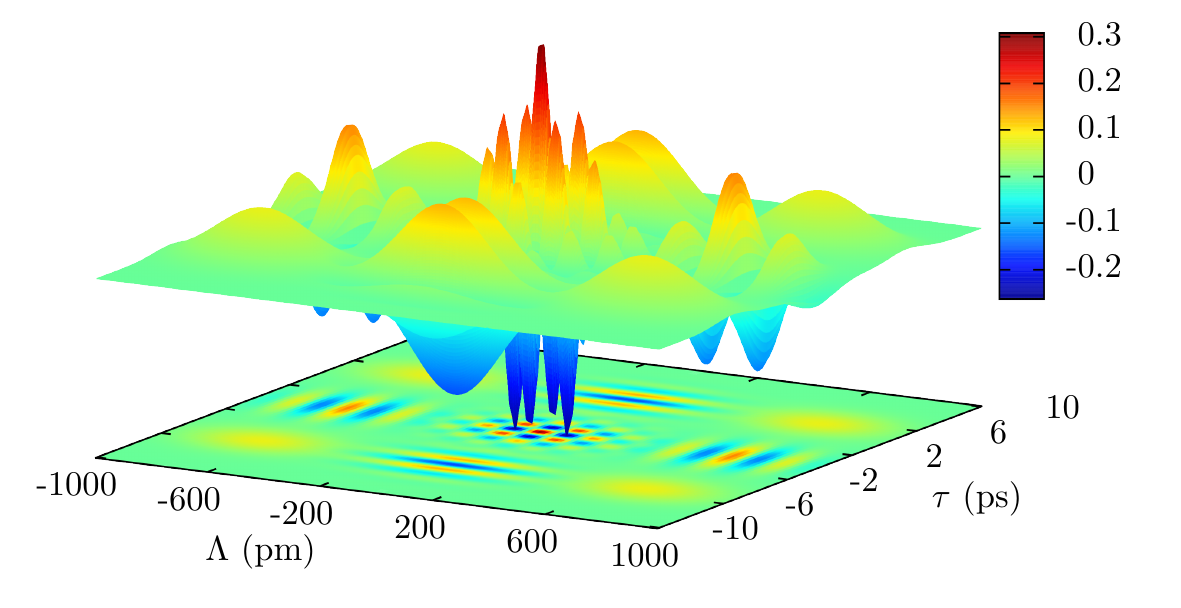}}
\end{minipage}
\caption{(Color online) (a) Pump illumination scheme to generate a compass state : two pairs of beams impinge onto the waveguide at two spots $z_a$ and $z_b$ equidistant from the center of the source, each pair consisting of two beams tilted symmetrically with respect to the degeneracy angle ($\theta_{a,b} = \theta_{\mathrm{deg}}\pm\delta\theta$). (b) Norm and (c) phase of the biphoton wavefunction of the corresponding cat state with $\delta\theta = 9.37\arcminute$ , $|z_a - z_b| = 1\milli\meter$ and for each beam $w_p = 200\micro\meter$, $\lambda_p = 775\nano\meter$, $\tau_p = 3.2\pico\second$ and $\theta = \theta_{\mathrm{deg}}$. (d) Corresponding Wigner function with $\Delta\Lambda = 1.37 \nano\meter$ and $\Delta\tau = 10.8\pico\second$. The $\Omega$ axis is given in units $\Lambda = \frac{8\pi c}{\omega_p^2}\Omega$. }
\label{fig:fourCat}
\end{figure*}

In order to clarify these concepts, we present the case study of a tranversally pumped semiconductor waveguide \cite{orieux2011JotOSoABOPEfficient, orieux2013PRLDirect} potentially having a great versatility in the control of the biphoton frequency correlations \cite{eckstein2014LPRHigh}. Indeed it has been shown that the pumping configuration allows one to modify the phase matching conditions by simply changing the spatial properties of the pump beam\cite{caillet2009JoMOsemiconductor,pefmmodeheckrlseriina2008PRAQuantum,walton2004PRAGeneration}, e.g. its waist or angle of incidence.  However,  more complex quantum state engineering has not been explored so far. 
\par  
The properties of the photon pair are described by the biphoton wavefunction which takes a complex form due to the constraints imposed by momentum and energy conservation. By neglecting group velocity dispersion (which is justified in the spectral range of the generated photon pairs) and in the narrow bandwidth limit for the pump beam, we can write the state of the pair under the following form:
\begin{equation}
\ket{\Psi} = \chi_\Gamma \iint \ud\omega_s \ud\omega_i f_+(\omega_s + \omega_i)f_-(\omega_s - \omega_i)\ket{\omega_s,\omega_i}
\end{equation}
where $\omega_{s,i}$  represent  the signal and idler frequencies, respectively,  and $\chi_\Gamma$ is a normalization constant\footnote{The overlap between the fields in the transverse and epitaxial directions of the waveguide is contained in this term. The variations of $\chi_\Gamma$ with the pump, signal and idler frequencies are negligible in the range of this study.}. The function $f_+$ corresponds to the spectrum of the pump beam reflecting the condition of energy conservation ($\omega_p = \omega_s + \omega_i \equiv \omega_+$). The phase of this function is affected by the presence of a microcavity formed by two Bragg mirrors as can be seen in the non flat phase profile in Fig.~\ref{fig:JSA_phase}.
The narrow band assumption allows us to consider the phase-matching dependent part of the biphoton state as a strictly antidiagonal function $f_-$: 
\begin{equation}
f_-(\omega_- \equiv \omega_s-\omega_i) = \int^{L/2}_{-L/2} \ud z \varphi(z) e^{i \frac{(\omega_s-\omega_i) z}{\bar{v}_g}},
\end{equation}
where $L$ is the length of the sample, $\bar{v}_g$ is the average group velocity of the signal and idler photons at frequency degeneracy and $\varphi(z) = \Phi(z) e^{-ik_{\mathrm{deg}}z}$. The function $\Phi(z)$ is the spatial profile of the pump beam and $k_{\mathrm{deg}} =\sin(\theta_ {\mathrm{deg}})\dfrac{\omega_p^{(0)}}{ c} =\dfrac{(n_s - n_i)}{2}\dfrac{\omega_p^{(0)}}{ c}$, where $n_{s,i}$ corresponds to the signal and idler effective refractive indices at degeneracy, $\omega_p^{(0)}$ is the central frequency of the pump beam and $\theta_{\mathrm{deg}}$ is the pumping incidence angle for which degeneracy occurs.

If the dimensions of the waveguide are large with respect to the pump waist, i.e. the limit where $L\rightarrow \infty$, $f_-$ can be approximated by the Fourier Transform of the spatial profile of the pump beam:
\begin{equation}
f_- (\omega_-) \approx \tilde{\varphi}\left(\frac{\omega_-}{\bar{v}_g}\right).
\end{equation}
We start by considering the situation depicted in Fig.~\ref{fig:source} where a gaussian pump beam with waist $w_p$ impinges onto the source at an angle $\theta$ and position $z_0$. The field distribution along the axis $z$ is 
$\Phi(z) \propto e^{-{(z-z_0)^2 \cos^2{\theta}}/{w_p^2}} e^{ i (k\sin\theta) z}$ and therefore $f_-$ reads:

\begin{equation}
f_-(\omega_-)\propto e^{i\omega_- \tau_0} e^{-\tfrac{\left(\omega_- - \omega_-^{(0)}\right)^2 }{\Delta\omega^2}}
\label{eq:coherentState}
\end{equation}
with $\tau_0 = z_0/\bar{v}_g$, $\Delta\omega = \bar{v}_g/2 w_p$ and $\omega_-^{(0)} = (k\sin\theta-k_{\mathrm{deg}})\bar{v}_g \approx \delta\theta \bar{v}_g \, \omega_p / c$
\footnote{The transverse size of the pump beam will change with the incidence angle. However the range of angles considered here is small enough to approximate $\sin{\theta}\rightarrow \theta$ and $\cos{\theta}\rightarrow 1$}. Fig.~\ref{fig:JSA_norm} and Fig.~\ref{fig:JSA_phase} represent the biphoton wavefunction norm and phase distribution numerically simulated for a pump impinging at the degeneracy angle $\theta_{\mathrm{deg}}$ at $z_0 = 0$ with waist $w_p = 200\micro\meter \ll L = 2\milli\meter$, central wavelength $\lambda_p = 775\nano\meter$ and pulse duration $\tau_p = 3.2\pico\second$ 
\footnote{\modif{This choice of parameters correspond to typical experimental conditions for the device illustrating our method. In particular, the generation of frequency uncorrelated biphoton states is achievable by modifying the size pump beam so that $f_+$ and $f_-$ have similar widths. Note that the technique remains valid for other pumping regime since the pulse duration only impacts the positively correlated part $f_+$ of the biphoton wavefunction.}}.

Instead of the complex-valued biphoton wavefunction, a more convenient representation is given by chronocyclic Wigner functions \modif{which are the time-energy analogs of the phase-space Wigner functions}. With this approach, time and energy properties of the single photons of the pairs are illustrated with real quantities\cite{brecht2013PRACharacterizing,sanchez-lozano2012JoOrelationship}, the same way pulses of light are depicted in the domain of ultrafast optics\cite{paye1992chronocyclic}. In this work, we describe, instead of the properties of the isolated photons of the pair, their correlations along the antidiagonal part $f_-$ of the biphoton wavefunction. \modif{This represents the quasi-probability distribution of the biphoton as a function of the detuning $\Omega$ between signal and idler frequencies and the time delay conjugated variable $\tau$.} The corresponding Wigner function $W_-$ is given by:
\begin{equation}\label{eq:wigner}
W_-(\tau, \Omega)=\int_{-\infty}^{\infty}\ud\omega_-f_-(\Omega-\omega_-)f_-^*(\Omega+\omega_-)e^{i2\tau \omega_-}. 
\end{equation}
Using the expression obtained in \eqref{eq:coherentState}, we see that this corresponds to a gaussian Wigner function centered at point $\tau=\tau_0$, $\Omega = \omega_-^{(0)}$, of widths $\Delta\Omega = 1/\Delta\tau = \Delta\omega = \bar{v}_g/2 w_p$ that is equivalent to the representation of a coherent state (see Fig.~\ref{fig:simpleWigner}). Thus, in this situation, shifting the pumping spot $z_0$ is equivalent to realizing displacements in the $\tau$ axis of the phase space while changing the angle of incidence $\theta$ of the pump beam corresponds to shifting the state along the $\Omega$ axis. 

More complex states can be obtained by engineering the pump beam. Indeed two identical beams impinging at $z_a$ and $z_b$ will generate a superposition of two coherent states displaced along the $\tau$ axis. In the limit 
$|z_a - z_b| \gg 2 w_p$ such states are almost orthogonal, representing a superposition of two distinct quasi-classical states (Schr\"odinger cat-like states). An analog superposition is obtained along axis $\Omega$ by using 2 different angles of incidence $\theta_a$ and $\theta_b$, impinging at the same point $z_0$. Quasi-orthogonality is obtained for $|\theta_a - \theta_b| \gg \frac{c}{2\omega_p w_p}$. 
We can generalize these Schr\"odinger cat-like states using more complex  configurations of pump beams. As an illustration, we choose a compass state (see Fig.~\ref{fig:fourCat}), a superposition of four coherent states presenting interesting applications in quantum metrology, as pointed out in \cite{zurek2001NLSub,toscano2006PRASub}. In order to obtain such state, a set of 4 different pump beam configurations  is required: 2 pairs of beams impinging at 2 different points separated by a distance $\Delta z$, each pair consisting in 2 beams tilted symmetrically with respect to the degeneracy angle as shown in Fig.~\ref{fig:fourLegsCat}.

The procedure detailed above to generate quantum states with different properties and applications can be generalized, leading to the creation of arbitrary continuous variables quantum states, since any state can be constructed from the over-complete basis of coherent states. 

\begin{figure}[t!]
\centering
\includegraphics[width=0.49\textwidth]{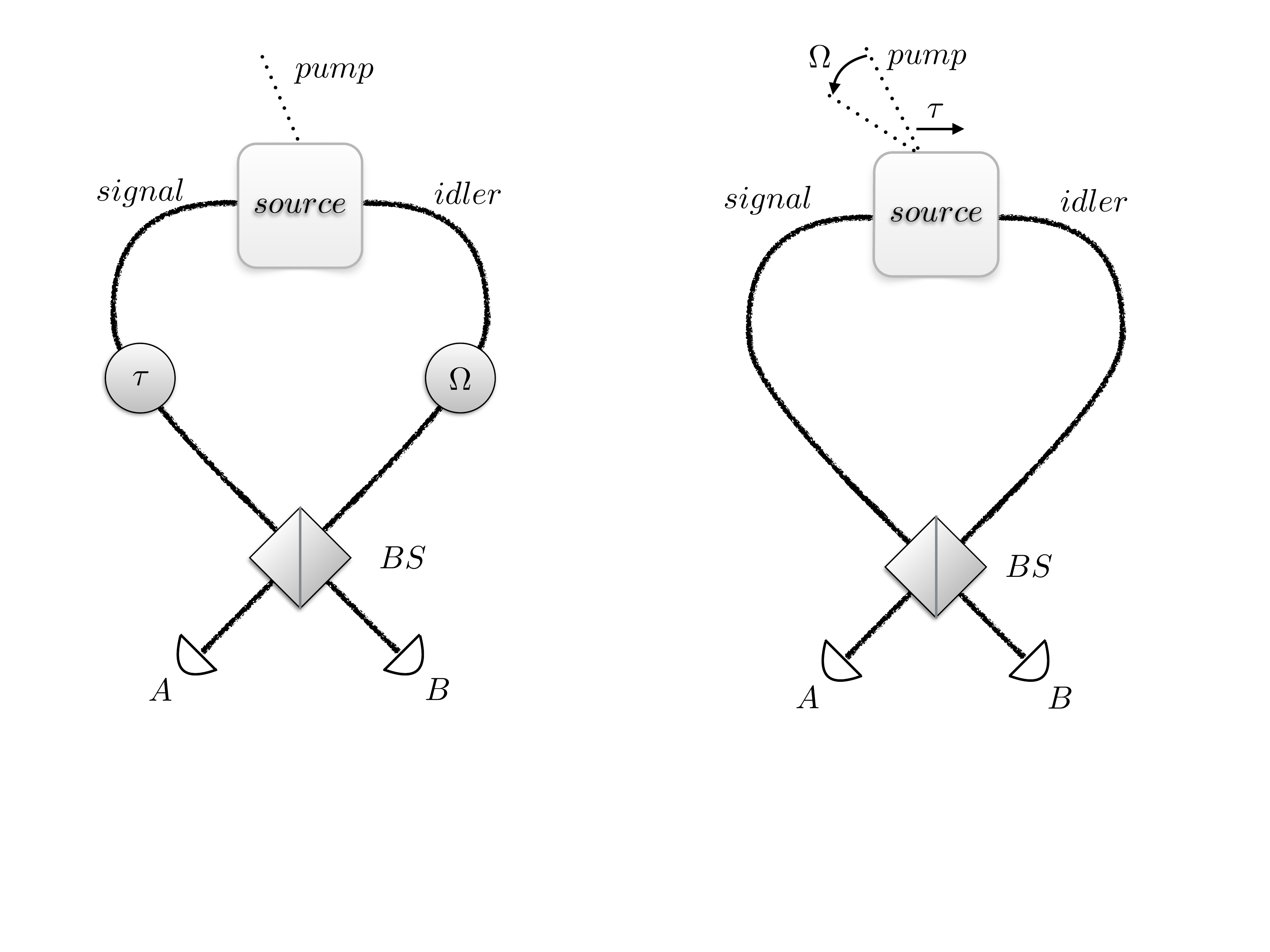}
\caption{Two possible strategies to realize the modified HOM experiment leading to the measurement of the Wigner function. Left: displacements in time and frequency are realized in each arm of the interferometer after the production of the photon pair. Right: in a completely analogous set-up, displacements in time and frequency are realized by pump engineering. While modifying the incidence angle of the pump beam, and consequently, the phase matching condition, corresponds to displacements in frequency, modifying the pump's incidence spot corresponds to displacements in time.\label{Fig3} }
\end{figure}

We now discuss how pump engineering can be used to characterize arbitrary CV states using the direct measurement of the Wigner function in all points of phase space. In \cite{douce2013SRDirect}, it was shown how a generalization of the HOM experiment leads to the direct measurement the Wigner function in the $\Omega$, $\tau$ phase space. Such generalization consists of considering displacements in the frequency degree of freedom of the photons,  in addition to time displacements. Of course, displacing either one of these parameters  modifies the distinguishability between the photons in each arm of the HOM interferometer, as depicted in Figure \ref{Fig3}. Time displacements can be realized relatively straightforwardly by simply modifying the optical path in each arm of the HOM interferometer. However, the frequency displacements required to reconstruct the Wigner function are quite broad with respect to the performances of currently available optical modulators \cite{olislager2012NJoPImplementing}.
Pump engineering provides an alternative solution to realize both time and frequency displacements, dramatically simplifying the direct measurement of the Wigner function and the CV state characterization.

As discussed previously, modification of the pump beam's incidence angle corresponds to displacing the central frequency of the symmetric part of the wave function associated to the  photon pairs, while modification on the incidence point corresponds to time displacements. Using these ingredients,  one can devise a procedure for the complete Wigner function measurement as follows:  in a first step, an initial state is engineered. Such state is the one to be characterized. Running the HOM experiment with no frequency or time displacement leads directly to the value of $W_-(0,0)$, the Wigner function at the origin \cite{douce2013SRDirect}. Then, tilting  the incident pump beams by a given amount, and repeating the same HOM experiment is equivalent to displacing in frequency the original state and measuring its Wigner function, leading to the value of $W_-(\Omega, 0)$. Analogously, displacing the pump beam in the $z$ axis and repeating the HOM strategy leads to the value of $W_-(0,\tau)$. It is clear that, by combining different tilting angles and displacements, one can obtain an arbitrary point of the Wigner function and reconstruct $W_-(\Omega, \tau)$ for all the values of $\Omega$ and $\tau$. In order to characterize the quantum state, the magnitude of the displacements in both axis of phase space should cover the region where $|f_-(\omega_-)|^2$ has a significant value. This corresponds to realizing the pump's angular displacements in an interval $\Delta \theta \approx \theta_a - \theta_b = 18.7\arcminute$, while its impinging position is displaced of $\Delta z \approx z_a - z_b = 1 \milli\meter$ (see Fig.~\ref{fig:fourCat}). Notice that, as shown in \cite{ou1988PRLObservation} the proposed strategy presents the advantage of not being limited by the detector's response time for measuring highly oscillating fringes or phase space structures associated with sub-Planck scales \cite{zurek2001NLSub}. 
\modif{Note that the resolution required for displacements along the $z$-axis is easily achieved; as far as the angular displacement is concerned, since $\Delta\theta$ is of the order of $\unit{2.7}{\milli\radian}$, a resolution of $\unit{100}{\micro\radian}$ is sufficient to resolve the fringes. This is also achievable with thermal stabilization and stable mechanical mountings.}

One may argue that modifying the pump, in reality, modifies the state to be measured instead of the measuring apparatus that is probing different points of the phase space; but this is common practice in quantum measurement strategies,  where the modification of the settings of the measurement apparatus is formally equivalent to that of the state-to-be-measured. This equivalency has been previously used, for instance in the context of cavity quantum electrodynamics in \cite{deleglise2008NLReconstruction}. There, the Wigner function is directly measured  using a Rydberg atom interacting dispersively for a fixed time interval with the field of a high quality micro wave cavity. The set-up is kept the same for all points of phase space and the quantum state of the field in the cavity to be measured is displaced, and consequently modified, through the application of a coherent field. Also in \cite{tischler2015Measurement}, the use of a monochromatic pump generating pairs of photons through SPDC is combined with modification of the pump frequency (or shifts in the temperature of the crystal) to circumvent the difficulty of broad frequency displacements. In the present case, changing the pump spatial configuration is equivalent to displacing the state to be measured.

In conclusion, we have shown how spatial pump engineering can be used to generate arbitrary CV frequency states of a photon pair and directly characterize it through a HOM-like experiment. The combination of the variety of pumping geometries with that of the possible design of integrated photonic circuits \cite{obrien2009NPPhotonic} will allow the realization of complex and versatile quantum photonic chips \cite{walton2004PRAGeneration}. The generalization of this technique to prepare quantum states with higher photon numbers, as well as to simulate the dynamics of CV states and the application of different quantum operations to it is an interesting open perspective that will be investigated.

%

\end{document}